\documentclass[%
 reprint,
 amsmath,amssymb,
 aps,
]{revtex4-1}

\usepackage{hyperref}
\usepackage{graphicx}%
\usepackage{dcolumn}%
\usepackage{bm}%
\usepackage{xcolor}  %
\usepackage{appendix}
\usepackage{upgreek}
\usepackage{soul} %

\begin{document}

\newcommand{\trm}[1]{\textrm{#1}}
\newcommand{\uuuline}[1]{\underline{\underline{\underline{#1}}}}
\newcommand{\prens}[1]{\left(#1\right)}
\newcommand{\um}{\upmu\trm{m}}
\newcommand{\nm}{\trm{nm}}
\newcommand{\cjs}[1]{\textcolor{blue}{#1}}
\newcommand{\ydd}[1]{\textcolor{red}{#1}}
\newcommand{\tbmu}{t_{\trm{b}\upmu}}

\preprint{APS/123-QED}

\title{Piezoelectric transduction of a wavelength-scale mechanical waveguide}%

\author{Yanni D. Dahmani}
\thanks{These authors contributed equally to this work.}
\author{Christopher J. Sarabalis}
\thanks{These authors contributed equally to this work.}
\author{Wentao Jiang}
\author{Felix M. Mayor}
\affiliation{%
 Department of Applied Physics and Ginzton Laboratory, Stanford University\\
 348 Via Pueblo Mall, Stanford, California 94305, USA
}
\author{Amir H. Safavi-Naeini}
\email{safavi@stanford.edu}
\affiliation{%
 Department of Applied Physics and Ginzton Laboratory, Stanford University\\
 348 Via Pueblo Mall, Stanford, California 94305, USA
}%

\date{\today}%

\begin{abstract}
    We present a piezoelectric transducer in thin-film lithium niobate that converts a 1.7~GHz microwave signal to a mechanical wave in a single mode of a 1~micron-wide waveguide. We measure a -12~dB conversion efficiency that is limited by material loss.  The design method we employ is widely applicable to the transduction of wavelength-scale structures used in emerging phononic circuits like those at the heart of many optomechanical microwave-to-optical converters.
\end{abstract}

\pacs{Valid PACS appear here}%
\maketitle

\section{\label{sec:level1}Introduction}
Mechanical waves play important roles in classical and quantum microwave systems from low-power, passive filtering and signal processing~\cite{Lakin1993,Manzaneque2017a,cleland2019} to compact resonators for state storage~\cite{Satzinger2018,Arrangoiz-Arriola2019,Gustafsson207} and low-loss channels for routing and delays~\cite{Fang2016,Patel2018,Manzaneque2019}. In optical systems, phonons mediate interactions between photons enabling efficient and compact modulators~\cite{Fan2019,Liu2019,Tadesse2014c,Sohn2018b,Merklein2017,Poulton2013} and full-spectrum beam steering~\cite{Sarabalis2018} with promising routes to microwave-to-optical conversion for quantum state transfer~\cite{Vainsencher2016,Bochmann2013,Balram2017b,Forsch2018}. 

Emerging applications of on-chip mechanics has led to the development of phononic circuits that utilize components which have dimensions on the order of the mechanical wavelength~\cite{Olsson2009,Mohammadi2011,Fu2019}. Efficient and selective transducers of wavelength-scale waveguides are needed to fulfill the promise of on-chip phononics.  The modes of small waveguides have been selectively transduced with optical forces in cavity optomechanics~\cite{Patel2018,Fang2016} and stimulated Brillouin scattering~\cite{VanLaer2015a,Kittlaus2016,Kang2010a}. In these devices, an optical-frequency photon scatters, emitting a microwave-frequency phonon with \(10^5\) times less energy. Generating a large phonon flux requires impractically large optical power because of this difference in energy scales~\cite{Safavi-Naeini2018}. Capacitive transducers~\cite{Romero2019,VanLaer2018a} and piezoelectric transducers convert microwave photons to phonons of the same frequency and thus near unity power efficiency conversion can be realized.

Piezoelectric transducers have been used to excite wavelength-scale devices including nanobeam resonators~\cite{Vainsencher2016,Balram2017b,Forsch2018,Jiang2019} and waveguides~\cite{Liu2019,Siddiqui2018} in a host of material platforms including aluminum nitride, aluminum nitride on silicon, galium arsenide, and more recently lithium niobate (LN).
Some recent approaches to integrated acousto-optic modulation~\cite{Liu2019} and microwave-to-optical quantum conversion~\cite{Forsch2018,Jiang2019} have suffered primarily from inefficient microwave-to-mechanical conversion and thus can benefit from a more systematic, component-based approach to the design and characterization of efficient, single-mode transducers.
One of the difficulties in designing such a device is in coupling a many-wavelength-scale transducer to a waveguide with a much smaller cross-section.  As discussed in Section~\ref{sec:design}, the area of these transducers scales inversely with the strength of the piezoelectric coupling which is captured in a parameter called $k^2_\text{eff}$.  In suspended LN, resonators~\cite{Yang2018} and delay line transducers~\cite{Sarabalis2019TheTransducers,Manzaneque2017a} have been demonstrated with \(k_\trm{eff}^2\) on the order of 10\%. Compared to previous work in aluminum nitride and galium arsenide where \(k_\trm{eff}^2\) is on the order of 1\%~\cite{Wu2019}, suspended LN opens opportunities to make more compact transducers enabling new approaches to their design.

We present an approach to designing efficient transducers of wavelength-scale guided modes and demonstrate a single-mode transducer for GHz-frequency, horizontal shear (SH) waves of a $1~\um$-wide rectangular waveguide in suspended, thin-film LN. We restrict our attention to narrow transducers that stay in the low-mode-density limit where normal-mode analysis is practicable, thereby avoiding the challenges of beam forming and horn design~\cite{Siddiqui2018}. In this work, we find that engineering a transducer requires careful consideration of the trade-offs between device footprint, bandwidth, and efficiency. Modeling our transducers with the finite element method (FEM)~\cite{comsol}, we decompose the radiated mechanical waves in the basis of modes of the $1~\um$-wide waveguide to find that 96\% of the emitted mechanical power  is into the SH0 mode.  We use these models to understand the limitations on the efficiency and bandwidth imposed by loss. We demonstrate compact $260~\um^2$ transducers with \(10~\um\) long horns that exhibit an electrical-to-SH0-mode transmission coefficient $|\tbmu|^{2}$ of $-12~\trm{dB}$ limited primarily by material loss.  Finally we characterize the SH0 mode at $1.7~\trm{GHz}$, finding a group velocity of $4.0\times 10^3~\trm{m}/\trm{s}$ and an attenuation coefficient of  \(\alpha=6.8~\trm{dB/mm}\) in close agreement with material losses in other work on LN films~\cite{Sarabalis2019TheTransducers,Vidal-alvarez2017a,Yang2018}.

\begin{figure*}[t]
    \centering
    \includegraphics[width=1\textwidth]{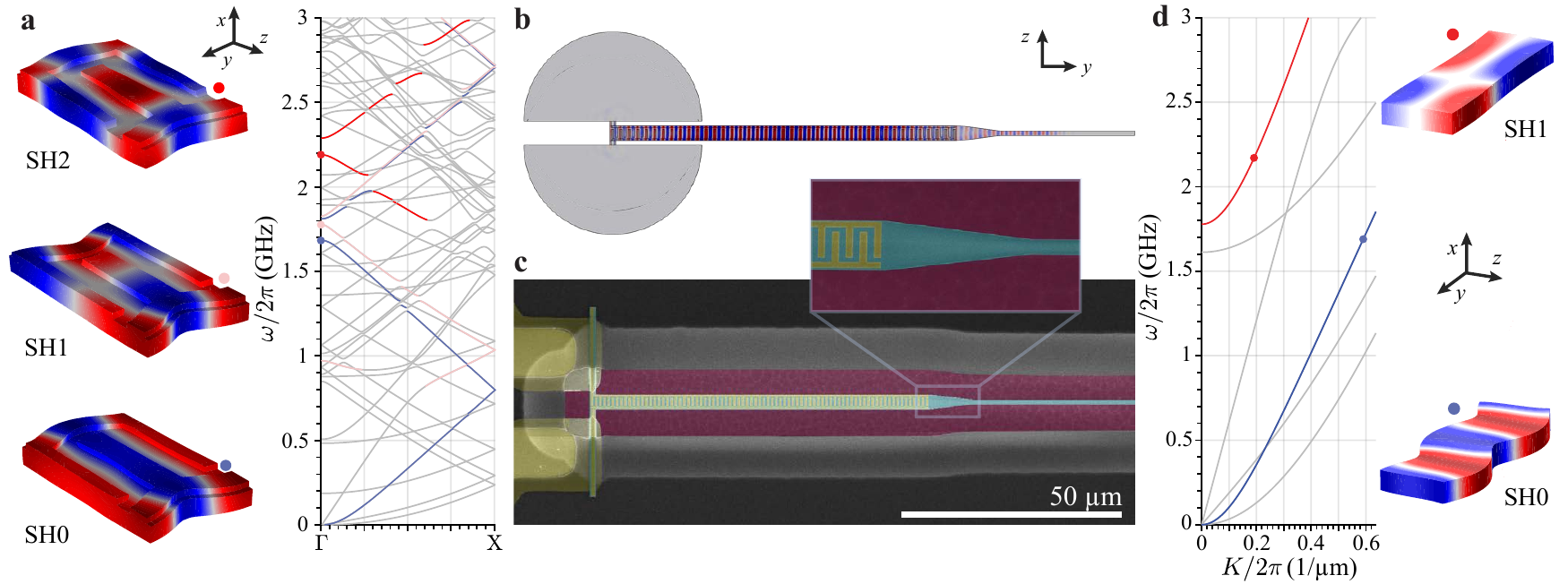}
    \caption{Suspended transducers patterned in \(300~\nm\)-thick, X-cut LN on silicon are designed to excite the SH0 mode of a \(1~\um\)-wide waveguide at \(1.7~\trm{GHz}\).  They are composed of a \(3.4~\um\) wide, \(100~\trm{nm}\) thick aluminum IDT and a \(10~\um\) long linear horn.  In the false color SEM \textbf{c}, LN is blue, aluminum is yellow, and the XeF\(_2\) release etch front is burgundy. FEM analysis \textbf{b} shows the horn scatters the SH0 mode of the IDT efficiently into the SH0 mode of the \(1~\um\)-wide waveguide. Bands and Bloch functions of the IDT and waveguide which constitute the asymptotic state of the horn are plotted at left \textbf{a} and right \textbf{d}, respectively. Waves propagate along \(y\).}
    \label{fig:Fig1}
\end{figure*}

\section{Transducer Area}
\label{sec:design}

We begin by considering the footprint $A$ needed to impedance match to a $50~\Omega$ transmission line.
Even without detailed knowledge of an IDT's modes, its footprint is constrained by the piezoelectric coupling coefficient \(k_\trm{eff}^2\)~\cite{SarabalisInPrep,Sarabalis2019TheTransducers}
\begin{eqnarray}
    A &=& \frac{\pi}{4}\frac{1}{\omega_0^2 c_\trm{s} k_\trm{eff}^2} \int\trm{d}\omega\, G\prens{\omega}\\
    &=& \frac{\pi}{8}^2\frac{G_0}{\omega_0^2 c_\trm{s}}\frac{\gamma}{k_\trm{eff}^2} ~~ (\text{for Lorentzian}~G(\omega))
    \label{eq:ksqInt}
\end{eqnarray} 
Here \(G\) is the conductance of the IDT (the real part of the admittance $Y$), $c_\trm{s}$ is its capacitance per unit area, $\omega_0$ is its center frequency, and the integral is evaluated over an interval about \(\omega_{0}\).  From Equation~\ref{eq:ksqInt} we see that matching to $50~\Omega$ over a large bandwidth comes at the cost of footprint. 
Materials like LN with high \(\varepsilon k_\trm{eff}^2\), where \(\varepsilon\) is the dielectric permittivity, enable small transducers with large bandwidth. 

Our goal is to efficiently excite SH waves in a $300$~nm thick, suspended, X-cut LN slab using a transducer matched to a $50~\Omega$ transmission line over a bandwidth of \(\gamma = 2\pi\times 10~\trm{MHz}\). We choose a $\gamma$, taken to be the full-width-half-maximum of the conductance, that is larger than the limit imposed by material loss.
From FEM models of an IDT unit cell, we determine an \(a = 1.9~\um\) pitch IDT with fingers parallel to the extraordinary axis excites SH waves at $1.7~\trm{GHz}$ and has a capacitance density \(c_\trm{s} = 155~\upmu\trm{F}/\trm{m}^2\). 
We assume a Lorentzian lineshape for $G(\omega)$ and \(k_\trm{eff}^2\) of \(35\%\)~\cite{Kuznetsova2001} leading to an IDT footprint  of \(A=250~\um^2\). 
To go from our footprint estimate to the precise dimensions of the transducer, we first consider guided modes of the mechanical waveguide and transducer. 

\section{The modes of LN waveguides: Choosing IDT dimensions}
\label{sec:waveguide-mechanics}

A piezoelectric waveguide with continuous translational symmetry such as the rectangular waveguide in Figure~\ref{fig:Fig1}d supports a power-orthogonal basis of modes at each frequency \(\omega\). These modes solve an eigenvalue problem on a 2D cross-section of the waveguide in which the stress $\boldsymbol{\sigma}$ and velocity $\mathbf{v}$ fields of the theory of elasticity and the electrostatic potential $\Phi$ of electrostatics are coupled by the piezoelectric tensor $\mathbf{d}$. The modes $\left|\psi_m\right\rangle \equiv \prens{\boldsymbol{\sigma}_m, \mathbf{v}_m, \Phi_m}$, indexed by $m$, vary along the waveguide as $e^{i K_m y}$ for complex eigenvalue $K_m$. If $K_i \neq K_j^*$, modes $i$ and $j$ are power-orthogonal and satisfy
\begin{align}
    \left\langle \psi_i \right| \left. \psi_j \right\rangle &\equiv -\int\trm{d}\vec{S}\cdot \prens{\boldsymbol{\sigma}_i^* \mathbf{v}_j + \boldsymbol{\sigma}_j \mathbf{v}_i^* + i\omega \mathbf{D}_i^* \Phi_j - i\omega \mathbf{D}_j \Phi_i^*}\nonumber \\
    &= 0 
    \label{eq:inner-product}
\end{align}
forming an inner-product space in which band structures can be computed and scattering can be studied~\cite{Auld1990v2}. Here \(\mathbf{D} = -\boldsymbol{\varepsilon}\nabla\Phi + \mathbf{d} \boldsymbol{\sigma} \) is the electric displacement field, and we normalize our basis such that $\left\langle \psi_i \right| \left. \psi_j \right\rangle = \delta_{ij}$. For more detail on our choice of Fourier conventions and the relationship between the inner-product and power see Appendix~\ref{sec:A1}.

The wavelength-scale LN waveguide we are trying to excite is $300~\trm{nm}$ thick and 1~\(\um\) wide, and supports four modes between 0 and \(1.6~\trm{GHz}\): the Lamb (A0), the shear (SH0), the first excited Lamb (A1), and the longitudinal (S0) mode~\cite{Auld1990v1}. The band structure is plotted in Figure~\ref{fig:Fig1}d.  For X-cut LN with crystal Z transverse to the waveguide, the shear mode couples to the electric field component \(E_y\) by the largest component of the piezoelectric tensor \(d_\trm{YZY} = 68~\trm{pC}/\trm{N}\)~\cite{Weis1985,Andrushchak2009}. For this reason we focus our attention on the shear bands.

\begin{figure}[h!]
    \centering
    \includegraphics[width=1\columnwidth]{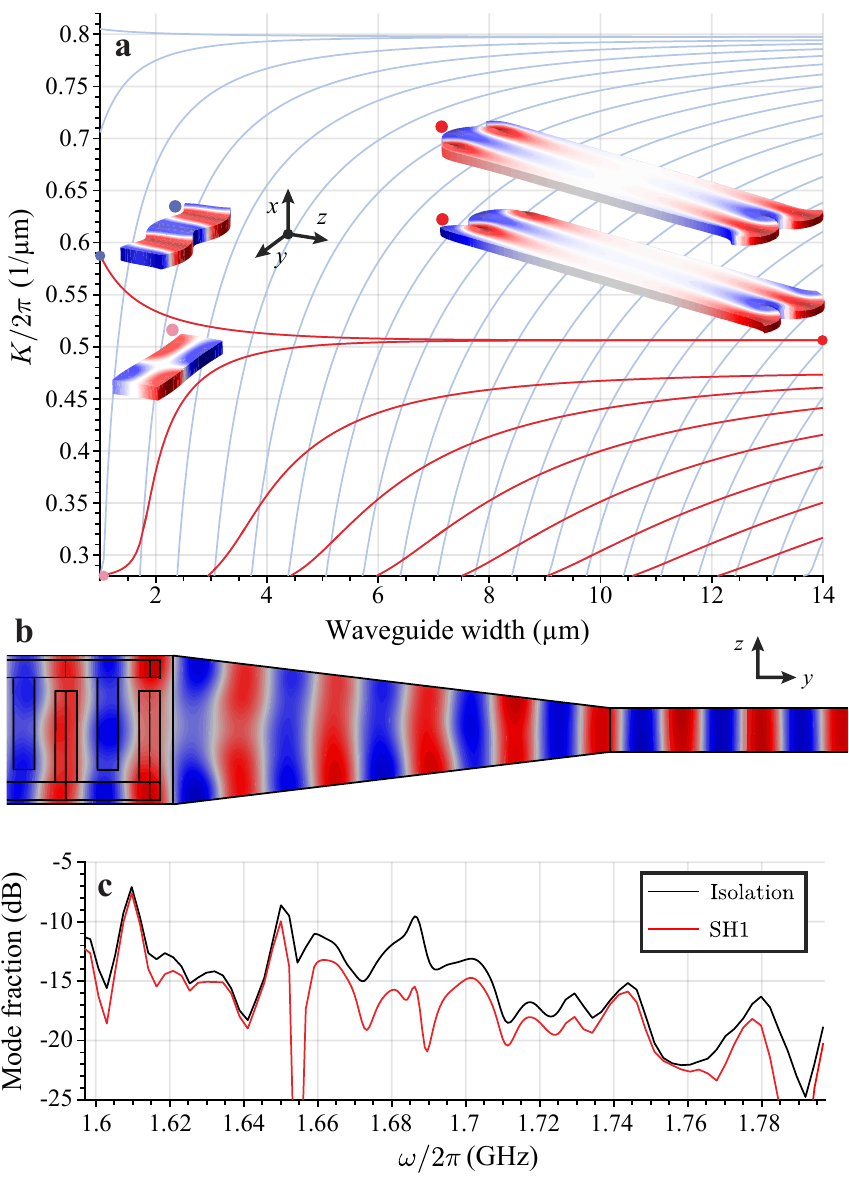}
    \caption{\textbf{a.} The number of modes supported by a suspended LN waveguide increases with width. 
    The SH0 and SH1 modes ($u_z$ plotted) limit to degenerate antisymmetric and symmetric edge supermodes, respectively. Below \(5~\um\) the SH waves (red) are well-resolved. The lamb waves are plotted in light blue. \textbf{b.} The linear horn scatters the SH0 mode of the \(3~\um\)-wide IDT efficiently into SH0 of the \(1~\um\)-wide output waveguide.  \textbf{c.} Decomposing the power in the waveguide we find that transduction of spurious modes, the isolation, is no greater than $-10$~dB away from the nodes in the conductance over a $200$~MHz bandwidth. The largest spurious component, the SH1 mode, is plotted in red. }
    \label{fig:densityOfStates}
\end{figure}

In these waveguides the number of modes increases with width and the first two SH modes become degenerate. For \(\omega = 2\pi\times 1.7~\trm{GHz}\), we numerically solve for the wavevectors \(K\) of the modes of a 2D cross section of the waveguide. At widths greater than 6~$\um$ (Figure~\ref{fig:densityOfStates}), the SH0 and SH1 modes localize to the edge, decaying exponentially away from the edges analogous to Rayleigh waves. These waves limit continuously to degenerate antisymmetric and symmetric supermodes of the edge states, respectively.

As shown in Figure~\ref{fig:densityOfStates}a, an adiabatic horn scatters the SH0 mode of a narrow waveguide to the antisymmetric supermode of a wide waveguide. Since the edge states are bound to the edge, their coupling to a wide IDT is limited. A wide IDT will excite modes with a significant amount of energy in the center of the waveguide. Mechanical power localized in the center of the IDT must either reflect off an adiabatic horn or scatter into modes other than the SH0 mode of the narrow waveguide. Therefore, for a wide IDT to efficiently excite the SH0 mode of the waveguide, the horn would need to mediate a non-adiabatic transition.

We choose $3.4~\um$ for the width of the IDT so that an adiabatically tapered horn can efficiently scatter the transduced mode into the \(1~\um\) waveguide. The narrow IDT allows us to simplify the design, make full use of the width of the transducer, spectrally resolve the SH0 and SH1 modes, and keep spurious shear modes in cutoff. We design a \(3.4~\um\)-wide transducer and a horn from \(3.4~\um\) to \(1~\um\) that efficiently scatters the output of the IDT into a single mode of the waveguide.

\section{Designing the IDT and Horn}
\label{sec:IDT}

From our footprint estimate in Section~\ref{sec:design}, we expect a \(3.4~\um\times 74~\um\) IDT to match to 50~\(\Omega\) corresponding to a \(N=39\) finger-pair transducer.  As we will see in simulation and experiment, material loss dominates the response at this length, reducing transmission from microwaves in the $50~\Omega$ line to phonons in the SH0 mode of the waveguide.  We focus our attention on a $40$ finger-pair IDT and, after discussing its performance, analyze the relationship between the transmission $\tbmu$ and $N$. 

Our waveguide transducer consists of three cascaded components, a $3.4~\um$-wide transducer, a $1~\um$-wide output waveguide with modes analyzed in Section~\ref{sec:waveguide-mechanics}, and a linear taper which connects the two.  A good transducer has a conductance reaching $20~\trm{mS}$ (matching to $50~\Omega$) and radiates efficiently into a single-mode of the output waveguide.

By orienting the IDT perpendicular to the crystal $\trm{Z}$ axis, we can leverage the $d_\trm{YZY}$ component of the piezoelectric tensor to efficiently excite the SH waves of the slab.  The $100~\trm{nm}$ aluminum electrodes mechanically and, for high \(k_\trm{eff}^2\) devices~\cite{Hashimoto2000a,Sarabalis2019TheTransducers}, electrically load the free slab of Section~\ref{sec:waveguide-mechanics}. The band diagrams and mode plots shown in Figure~\ref{fig:Fig1}a take this loading into account. The $1.92~\um$-pitch IDT has a duty cycle of 50\% with fingers that end $300~\trm{nm}$ away from the $400~\trm{nm}$ wide bus wires that run along the edges of the waveguide.  The $1.7$ and $2.2~\trm{GHz}$ SH0 and SH2 \(\Gamma\)-point modes of the IDT are efficiently transduced as seen in the measured \(S_{11}\) and conductance $G\prens{\omega}$ (Figure~\ref{fig:impulseResponse}b and c).  The symmetric $1.8~\trm{GHz}$ SH1 mode is weakly transduced due to symmetry.

\begin{figure}[h!]
    \centering
    \includegraphics[width=1\columnwidth]{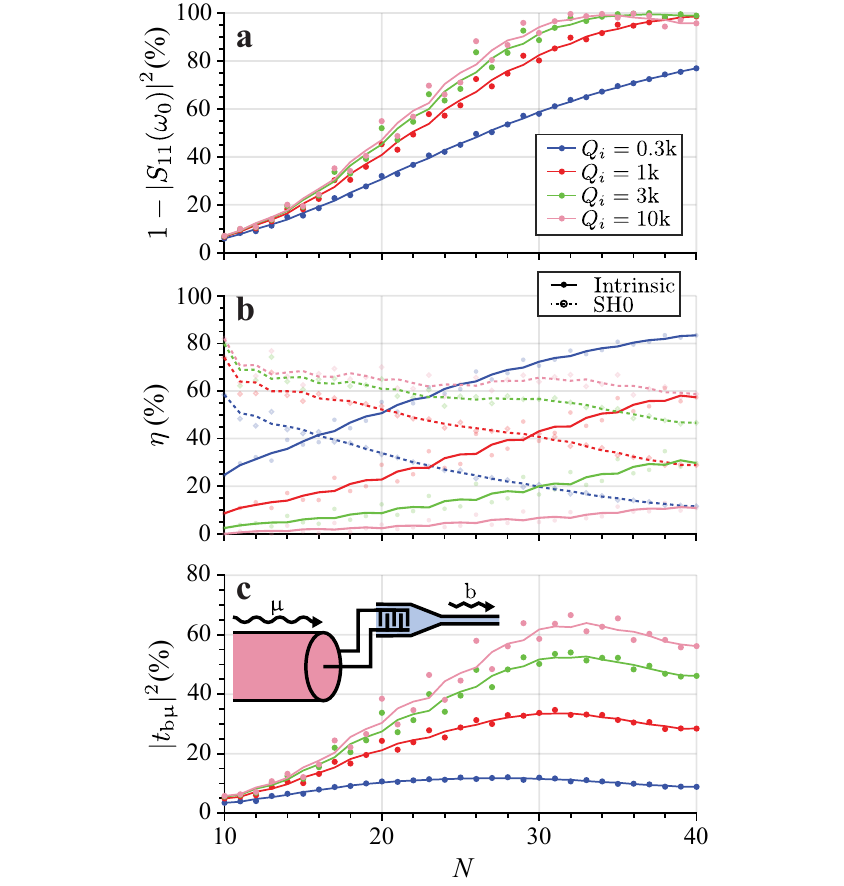}
    \caption{By computing the linear response \(Y\prens{\omega}\) and decomposition \(\lbrace a_m \rbrace\) we study the \(N\)-dependence of the transmission $\tbmu$ from a \(50~\Omega\) transmission line to the SH0 mode. \textbf{a.} As \(N\) increases so too does the peak conductance.  This decreases the microwave reflections \(S_{11}\).  \textbf{b.} Increasing \(N\) also decreases \(\gamma_\trm{e}\) and when \(N\) is large such that \(\gamma_\trm{i} \gg \gamma_\trm{e}\), material damping dominates the response.  This is seen in the fractions of the total power dissipated due to loss and due to transmission into SH0, \(\eta\).   \textbf{c.}  These competing effects lead to an optimal \(N\) for maximizing \(\left|\tbmu\right|^{2}\) which increases with \(Q_\trm{i}\).  Numerical results (points) are smoothed with a moving window average (curve) for clarity.}
    \label{fig:t_au}
\end{figure}

A 3D FEM model of the transducer, IDT, and taper, is used to compute both the linear response of the system and analyze the waves emitted. A solution on the full domain is shown in Figure~\ref{fig:Fig1}b.  A perfectly-matched layer bounds the domain. Reflections at the interface between the IDT and taper give rise to resonances which proportionally decrease the full-width-half-maximum and increase the peak height of $G\prens{\omega}$. In order to match the resonant response to our measurements we incorporate a uniform material loss tangent corresponding to \(Q_\trm{i} = 300\) and scale the piezoelectric tensor from its bulk values by $0.67$~\cite{Sarabalis2019TheTransducers}. The simulated and measured conductance of the SH0 mode is plotted in Figure~\ref{fig:impulseResponse}c for comparison. For the SH0 mode, the peak conductance of \(6.9~\trm{mS}\) and bandwidth (full-width-half-max) of \(7.3~\trm{MHz}\) are in good agreement with measurements, \(6.5~\trm{mS}\) and \(9.7~\trm{MHz}\).

The modes of the $1~\um$ waveguide studied in Section~\ref{sec:waveguide-mechanics} form a basis (see Appendix~\ref{sec:basis}) in which we can decompose the power radiated by the IDT and check that the transducer excites a single mode.
Given a solution $\left|\psi\right\rangle$, the coefficients $a_{m}$ are computed using the inner product defined in Equation~\ref{eq:inner-product}
\begin{equation}
    a_{m} = \left\langle\psi_m\right| \left. \psi\right\rangle
\end{equation}
such that
\begin{equation}
    \left|\psi\right\rangle = \sum_m a_m \left| \psi_m\right\rangle
\end{equation}
where \(\left|a_m\right|^2\) is the power in mode $m$. For each mode $m$ there's an associated backwards propagating mode $-m$, the pair of which form a piezoelectric port (see Appendix~\ref{sec:solutions-to-S}).

From the band diagram in Figure~\ref{fig:densityOfStates}a, we conclude that a horn consisting of a $10~\um$ linear taper functions approximately adiabatically. The power transmitted into the waveguide is transmitted into the SH0 mode with \(-10~\trm{dB}\) total power in spurious modes (labeled \emph{isolation} in Figure~\ref{fig:densityOfStates}c). Excluding nodes in the conductance (Figure~\ref{fig:impulseResponse}c), power in the next most strongly transduced mode (SH1) remains below $-15$~dB over $200~\trm{MHz}$.  This indicates that the $10~\um$ long linear taper efficiently scatters the SH0 mode of the $3.4~\um$ waveguide into the SH0 mode of the $1~\um$ waveguide. %

Of the total power dissipated by an $N = 40$ transducer, $2 G\prens{\omega} \left|V\prens{\omega}\right|^2$, \(11\%\) is emitted into the waveguide at the center frequency, $96\%$ of which is in the SH0 mode. Only $5\%$ is lost to clamping while the other \(84\%\) is lost to material damping.

We are interested in how this transducer behaves when coupled to a \(50~\Omega\) transmission line. In particular, we are interested in the transmission coefficient \(\tbmu\) from microwaves in the transmission line to phonons in the SH0 mode and how it varies with $N$. The relationship between the FEM analysis and components of the S-matrix like \(\tbmu\) is detailed in Appendix~\ref{sec:solutions-to-S}.
In our previous work~\cite{Sarabalis2019TheTransducers}, we've shown that as $N$ is increased, the extrinsic coupling $\gamma_\trm{e}$ drops as $N^{-3}$ for strongly resonant IDTs like those reported here.  In the high $N$-limit where \(\gamma_\trm{e} \ll \gamma_i\), the bandwidth $\gamma$ approaches the material limit $\gamma_i = \omega_{0}/Q_\trm{i}$.  And while, in this limit, increasing $N$ continues to increase the conductance $G$ thus decreasing reflections from impedance mismatch (Figure~\ref{fig:t_au}a), this power is lost to damping. In this regime, increasing $N$ seems to improve $S_{11}$, but actually reduces the microwave-to-mechanical transmission. For this reason, \(\tbmu\) cannot be extracted from \(S_{11}\). In Section~\ref{sec:measurements}, \(\tbmu\) is instead determined from a measurement of \(S_{21}\). In Figure~\ref{fig:t_au}b we sweep $N$ and compute the percentages of the total power dissipated by the IDT transmitted into the SH0 mode and lost to damping and see in Figure~\ref{fig:t_au}c that for \(Q_\trm{i} = 300\), \(\tbmu\) reaches \(12\%\) at \(N = 29\).

There are a few approaches to improve $\left|\tbmu\right|^2$ beyond \(12\%\). The most obvious is to improve the material parameters \(k_\trm{eff}^2\) and \(Q_\trm{i}\) (see Figure~\ref{fig:t_au}).
For applications in quantum science, operating at cryogenic temperatures will likely increase $Q_\trm{i}$ by suppressing thermally induced mechanical loss and ohmic dissipation in the electrodes. Another strategy is to reduce the reflection coefficient at the IDT-waveguide interface thus increasing \(\gamma_\trm{e}\) for a given $N$ and allowing us to make longer transducers before reaching loss-limits. Lastly we could diverge from the low width, low density of states design and employ wider waveguides, embracing the challenges of multi-mode design~\cite{Siddiqui2018}.

\begin{figure}[h!]
    \centering
    \includegraphics[width=1\columnwidth]{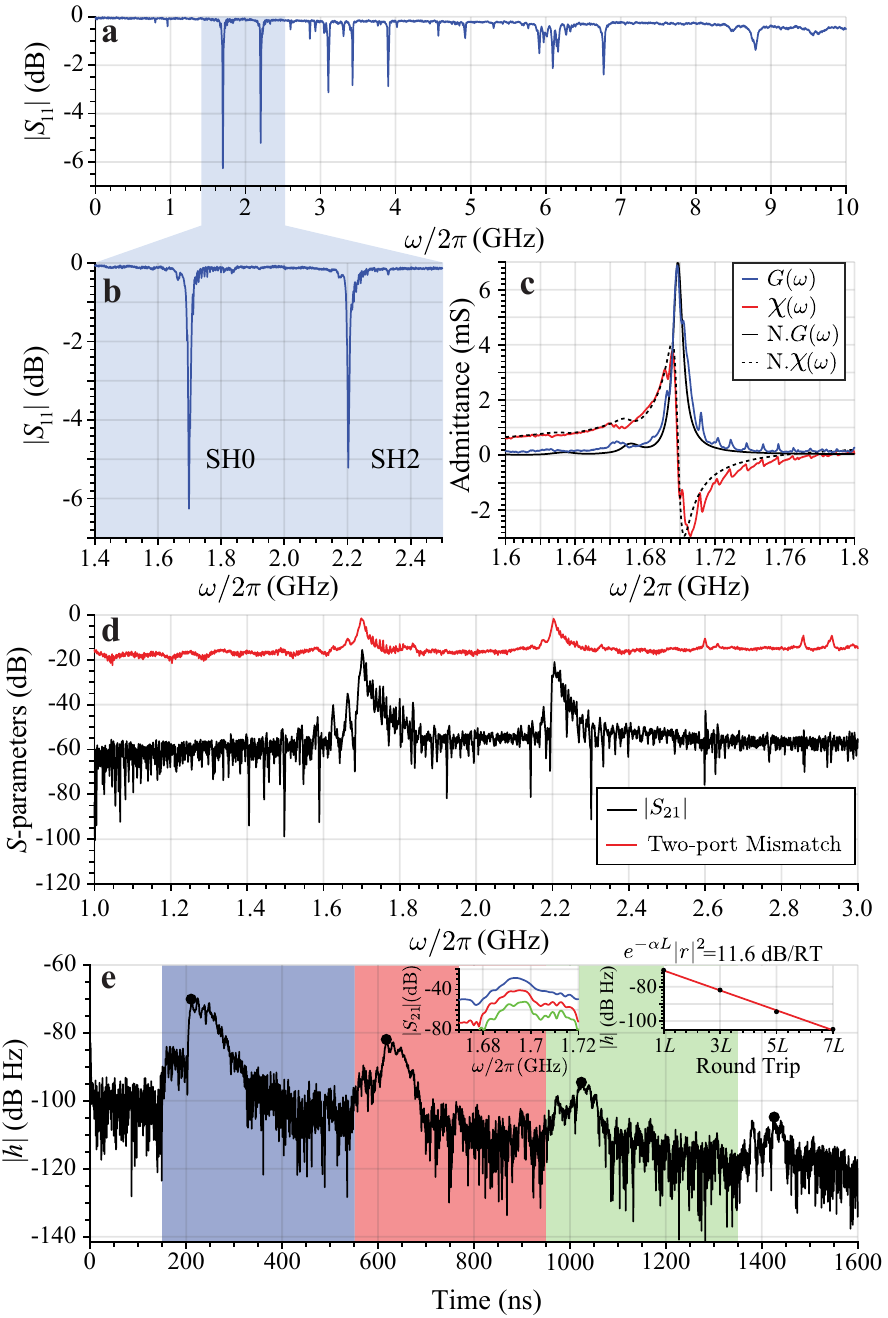}
    \caption{\textbf{a.} The \(\left|S_{11}\right|\) of an \(N=40\) transducer. \textbf{b.} The \(\left|S_{11}\right|\) restricted to the SH0 and SH2 responses. \textbf{c.} The measured conductance \(G\) and susceptance \(\chi\) of the SH0 mode are overlaid on FEM results. The SH2 mode is also strongly transduced with a peak conductance of \(5.6~\trm{mS}\) and full-width-half-max of \(8.2~\trm{MHz}\). \textbf{d.} The \(|S_{21}|\) of a delay line with a \(L=200~\um\) waveguide shows large SH0 and SH2 responses at \(1.7\) and \(2.2~\trm{GHz}\). For the SH2, of the \(-21.1~\trm{dB}\) insertion loss,  we attribute \(-1.7~\trm{dB}\) to impedance mismatch. The \(\left|S_{21}\right|\) of an ideal delay line with no insertion loss would equal the two-port mismatch,  \(1-|S_{11}|^{2}/2-|S_{22}|^{2}/2\) (see Appendix~\ref{sec:2portMismatch}). \textbf{e.} For \(L=800~\um\), the heights of the echoes in the impulse response are fit (inset) to extract a round trip loss of \(11.6~\trm{dB}\). The \(S_{21}\) is filtered in the time-domain with rectangular envelopes (shaded blue, red, and green) to compute the single, triple, and quintuple-transit $S_{21}$ plotted with corresponding colors (inset).  The peak single-transit response is used as an estimate of $\left|\tbmu\right|^2e^{-\alpha L/2}$.}
    
    \label{fig:impulseResponse}
\end{figure}

\section{Measurements}
\label{sec:measurements}

Starting with a \(500~\trm{nm}\)-thick film of LN on a \(500~\um\)-thick silicon substrate, the film is thinned to \(300~\trm{nm}\) by argon milling before patterning an HSQ mask with e-beam lithography to define the waveguides.  The mask is transferred to the LN by angled argon milling~\cite{Jiang2019}. We then perform an acid clean to remove resputtered, amorphous LN. We deposit \(100~\trm{nm}\) of Al for electrodes and \(200~\trm{nm}\) Al for contact pads by e-beam lithography and photolithography, respectively; metal evaporation; and liftoff. Finally we release the structures with a masked \(\trm{XeF}_{2}\) dry etch.

The \(S\)-parameters of the transducers are measured with a vector network analyzer (Rhode \& Schwarz ZNB20) on a probe station calibrated to move the reference plane to the tips of the probes (GGB nickel 40A).
Several modes below $10$~GHz are strongly transduced as seen in the \(S_{11}\) plotted in Figure~\ref{fig:impulseResponse}.
The conductance \(G \equiv \trm{Re}Y\) and susceptance \(\chi \equiv \trm{Im}Y\) for the SH0 mode plotted in Figure~\ref{fig:impulseResponse}c match well with the overlaid simulated curve and \(\Gamma\)-point frequency of the IDT unit cell bands shown in Figure~\ref{fig:Fig1}a.
The peak conductance and full-width-half-max for the SH0 is \(6.5~\trm{mS}\) and \(9.7~\trm{MHz}\), inferred by Lorentzian fit.
We infer a static capacitance of $31$~fF by fit to the DC response of $\chi$ and use it along with the conductance fit by Equation~\ref{eq:ksqInt} to calculate a \(k_\trm{eff}^2\) of $15\%$.  This is decreased by the feedthrough capacitance of the contact pads.

For an \(L=200~\um\) long waveguide, the minimum loss through the structure containing a transducer, taper, waveguide, taper, and another transducer is \(|S_{21}| = -15.7~\trm{dB}\) at $\omega = 2\pi\times 1.7~\text{GHz}$ (Figure~\ref{fig:impulseResponse}d). We attribute \(-1.5~\trm{dB}\) to impedance mismatch from the measured $S_{11}$ and $S_{22}$.  %
Reflections at the IDT-waveguide interface resonantly enhance transmission through the waveguide which is why the peak \(\left|S_{21}\right|\) is significantly larger than \(|\tbmu|^{4}\).

The peak of \(\left|S_{21}\right|\) encompasses all loss mechanisms in the device including clamping, damping, excitation of spurious modes, and propagation loss and is affected by reflections and resonance. We can isolate the propagation loss \(\alpha\) and \(\tbmu\) by analyzing the time-domain impulse response \(h\prens{t}\), the inverse Fourier transform of \(S_{21}(\omega)\), plotted for a device with \(L = 800~\um\) (Figure~\ref{fig:impulseResponse}e).
The first pulse takes the shortest path through the device and is attenuated by \(|\tbmu|^{2}e^{-\alpha L/2}\). Each subsequent echo takes an additional round trip, is attenuated by \(|r|^{2}e^{-\alpha L}\), and delayed by \(2L/v_{g}=4.0\times10^{2}~\trm{ns}\). We fit \(|r|^{2}e^{-\alpha L}=-11.6~\trm{dB}\) from the peaks in Figure~\ref{fig:impulseResponse}e and transform the first pulse (blue) back to the frequency domain (inset) to find \(|\tbmu|^{2}e^{-\alpha L/2}=-28.6~\trm{dB}\).
More detail is provided in the Appendix~\ref{sec:deembed}.

The single-transit and round-trip loss are two constraints on three unknown quantities: \(|\tbmu|^{2}\), \(|r|^{2}\), and \(\alpha\). By sweeping the length of the device, all three parameters can be determined independently. In lieu of a length sweep, we ignore scattering into other modes and assume \(|\tbmu|^{2}+|r|^{2}=1\) at the IDT-waveguide interface to find a \(|\tbmu|^{2}\) of \(7.0\%\) (comparable to the simulated value of $8.9\%$ for \(N=40\)), an \(|r|^{2}\) of \(93\%\), and an \(\alpha\) of \(6.8~\trm{dB/mm}\).  If an additional loss channel of \(10\%\) (\(50\%\)) is added, \(|\tbmu|^{2}\) decreases to \(6.6\%\) (\(4.9\%\)) while \(\alpha\) decreases to \(6.3~\trm{dB/mm}\) (\(2.9~\trm{dB/mm}\)). Given the measured group velocity of \(v_{g}=4.0\times 10^3~\trm{m/s}\), this \(\alpha\) corresponds to a quality factor \(Q=\omega_{0}/\alpha v_{g}\) of $1700$ in the waveguide and an \(f_{0}Q\) of \(2.9\times10^{12}\) which is comparable to our previous work in multi-moded, high frequency delay lines with an \(f_{0}Q\) of \(4.6\times10^{12}\)~\cite{Sarabalis2019TheTransducers}. We see an order of magnitude improvement over delay lines in suspended LN employing the S0 mode at $350$~MHz where \(f_{0}Q = 0.45\times10^{12}\)~\cite{Vidal-alvarez2017a}. Resonators using antisymmetric thickness modes exhibit \(f_{0}Q\) products over twice as large (\(9.15\times10^{12}\))~\cite{Yang2018}.

\section{\label{sec:level1}Conclusions}

In the fields of optomechanics and Brillouin scattering, interaction rates increase as the mechanical mode decreases in size.
Whether photon-phonon interactions are being used to make compact acousto-optic modulators~\cite{Sohn2018b,Fan2019,Liu2019} or as an approach to quantum microwave-to-optical links~\cite{Bochmann2013,Vainsencher2016,Balram2017b,Forsch2018,Jiang2019}, there's a common need to efficiently and selectively transduce the modes of wavelength-scale mechanical structures.

Our hope is that insights from our approach to addressing this problem, here in suspended LN, can be generally applied to exciting wavelength-scale mechanical devices.  The explicit connection between footprint and bandwidth constrains the design space. We show that in the presence of loss, long IDTs have limited transmission coefficients which may improve in cryogenic environments. In the pursuit of efficient, broadband transducers at room temperature, our results motivate mitigating reflections at the waveguide-IDT interface and engineering horns for wider, multi-moded transducers. Efficient, single-mode transducers that are ready to incorporate into complex phononic networks fundamentally advance our control over mechanical degrees of freedom. 

\section*{Acknowledgements}
The authors would like to thank Rishi N. Patel, Patricio Arrangoiz-Arriola, and Timothy P. McKenna for useful discussions. This work was supported by a MURI grant from the U. S. Air Force Office of Scientific Research (Grant No. FA9550-17-1-0002), by a fellowship from the David and Lucille Packard foundation, and by the National Science Foundation through ECCS-1808100 and PHY-1820938. Part of this work was performed at the Stanford Nano Shared Facilities (SNSF), supported by the National Science Foundation under Grant No. ECCS-1542152, and the Stanford Nanofabrication Facility (SNF).

\appendix

\section{\label{sec:A1}Power Dissipation in the Fourier Domain}

The voltage \(V(t)\) can be expressed in the frequency domain
\begin{equation} 
    V(\omega) = \dfrac{1}{2\pi}\int_{-\infty}^{\infty}d\omega V(t)e^{i\omega t}
    \label{eq:fourier}
\end{equation}
and similarly for current \(I\) and admittance \(Y\). With our choice of Fourier convention, the power dissipated by an electrical element \(\mathcal{P}(t) = V(t)I(t)\) is the convolution of \(V(\omega)\) and \(I(\omega)\) in the frequency domain. Averaging this quantity in time extracts the DC component of the spectrum thus reducing the convolution to
\begin{equation}
    \langle\mathcal{P}\rangle = \int_{-\infty}^{\infty}d\omega V(\omega)I(-\omega)
\end{equation}

Since the voltage is a real-valued quantity, \(V^{*}(\omega)\) is equal to \(V(-\omega)\) (the same argument holds for $I(\omega)$ and $Y(\omega)$). Changing our limits of integration and using Ohm's law, \(I(\omega)=Y(\omega)V(\omega)\), we find

\begin{equation}
    \langle\mathcal{P}\rangle = \int_{0}^{\infty}d\omega Y^{*}(\omega)\left|V(\omega)\right|^{2}+Y(\omega)\left|V(\omega)\right|^{2}
\end{equation}

Since \(2\trm{Re}(Y)(\omega)=2G(\omega) = Y(\omega)+Y^{*}(\omega)\), the time average power dissipated by the electrodes is
\begin{equation}
    \mathcal{P}_{0} = 2\int_{0}^{\infty}d\omega G(\omega)\left|V(\omega)\right|^{2}.
    \label{eq:electrode_power}
\end{equation}

To determine the time-average power for a piezoelectric wave, we repeat the previous analysis starting from the instantaneous piezoelectric Poynting vector 
\begin{equation}
    \mathcal{P}_{\trm{piezo}}(t) = -\boldsymbol{\sigma}(t)\mathbf{v}(t)+\Phi(t)\partial_{t}\mathbf{D}(t)
\end{equation}
to find the time-average piezoelectric power
\begin{equation}
    \mathcal{P}_{\trm{piezo}} = -\int_{0}^{\infty}d\omega\int d\vec{S}\cdot(\boldsymbol{\sigma}^{*}\mathbf{v}+\boldsymbol{\sigma}\mathbf{v}^{*}+i\omega\mathbf{D}^{*}\Phi-i\omega\mathbf{D}\Phi^{*}).
\end{equation}
We compare this expression to the inner product in Equation~\ref{eq:inner-product} to confirm \(\left|a_m\right|^2\) is the time-average power in mode \(m\). We note that our time-average power differs from that of Auld's by a factor of \(1/4\)~\cite{Auld1990v2} resulting from differing Fourier conventions. All values reported are power ratios and thus factors of \(2\) from choices of convention drop out.

\section{Basis}
\label{sec:basis}

\begin{figure}[h!]
    \centering
    \includegraphics[width=1\columnwidth]{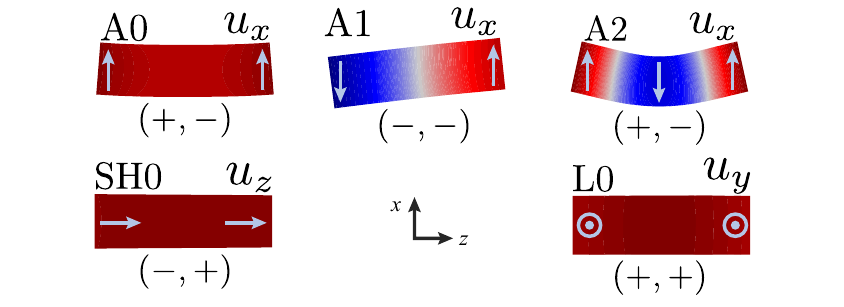}
    \caption{Modes of an LN waveguide at \(1.7~\trm{GHz}\). Color in these plots visualize the dominant displacement field. Light blue arrows show the direction of displacement.}
    \label{fig:modalBasis}
\end{figure}

Decomposition of the mechanical energy radiated into a waveguide is necessary for calculating transmission coefficients like \(\tbmu\) needed to characterize a phononic component.  For completeness we briefly describe the basis of propagating modes in a \(300~\nm\) thick, \(1~\um\)-wide, X-cut LN, rectangular waveguide. We categorize the five \(1.7~\trm{GHz}\) modes as Lamb (A), horizontal shear (SH), and longitudinal (S) modes which differ in their principal strains \(S_{xz}\), \(S_{yz}\), and \(S_{zz}\), respectively.  These modes are plotted in Figure~\ref{fig:modalBasis} along with their reflection symmetries \(\prens{\sigma_z,\sigma_x}\) where \(\prens{+,-}\), for example, means symmetric and antisymmetric with respect to reflection across the \(xy\) and \(yz\)-planes, respectively.

\section{Computing the S-matrix}
\label{sec:solutions-to-S}

In our FEM analysis, we are solving a set of inhomogenous equations describing the behavior of our piezoelectric device.  The drive term of these equations is a vector \(\prens{V, \mathbf{a}_-}^\top\) where \(V\) is the voltage across the leads of our transducer and \(\mathbf{a}_-\) is a vector of coefficients for the piezoelectric waves incident on the domain as defined by Equation~\ref{eq:inner-product}.

The solutions of these equations can be represented in matrix form
\begin{equation}
    \begin{pmatrix} I \\ \mathbf{a}_+ \end{pmatrix} = 
    \begin{pmatrix} Y & \mathbf{x}_1^\top \\ \mathbf{x}_2 & \trm{X} \end{pmatrix}
    \begin{pmatrix} V \\ \mathbf{a}_- \end{pmatrix}.
    \label{eq:G-matrix}
\end{equation}
The scalar \(Y\) is the admittance of the transducer.  For \(M\) modes, \(\mathbf{x}_1\) and \(\mathbf{x}_2\) are vectors with \(M\) components and \(\trm{X}\) is an \(M\times M\) matrix.  For the simulations reported here, the coefficients of \(\mathbf{a}_{-}\) are set to 0 and we solve for the first column of the matrix in Equation~\ref{eq:G-matrix} in terms of the input voltage \(V\).

In order to study how these transducers behave in phononic networks---for example, the two-port transmission devices we use to measure \(t_{\trm{b}\upmu}\)---we want to transform the matrix in Equation~\ref{eq:G-matrix} into a scattering matrix \(S\). To do so, we reexpress \(V\) and \(I\) in terms of the microwave amplitudes \(a_{\pm\upmu}\) which we abbreviate to \(a_\pm\) in this section
\begin{align}
    V &= \sqrt{\frac{Z_0}{2}}\prens{a_+ + a_-} \\
    I &= \frac{1}{\sqrt{2Z_0}}\prens{a_- - a_+}.
    \label{eq:microwave-amplitudes}
\end{align}
Here \(Z_0\) is impedance of the transmission line.
Like \(a_m\), the squares of the amplitudes \(a_+^2\) and \(a_-^2\) are the outward and inward-going, time-averaged power in the transmission line.
This is easily checked by computing the power into the microwave port
\begin{equation}
    V^*I + VI^* = \left|a_-\right|^2 - \left|a_+\right|^2.
\end{equation}
Substituting Equation~\ref{eq:microwave-amplitudes} into Equation~\ref{eq:G-matrix} and collecting terms we find the \(S\) matrix
\begin{equation}
    \begin{pmatrix} a_+ \\ \mathbf{a}_+ \end{pmatrix} = 
        \underbrace{\begin{pmatrix}
            \frac{Y_0 - Y}{Y_0 + Y} & -\frac{\sqrt{2Y_0}}{Y_0 + Y}\mathbf{x}_1^\top \\
            \frac{\sqrt{2Y_0}}{Y_0 + Y}\mathbf{x}_2 & \trm{X} + \frac{1}{Y_0 + Y}\mathbf{x}_2\mathbf{x}_1^\top
        \end{pmatrix}}_S
        \begin{pmatrix}a_- \\ \mathbf{a}_-\end{pmatrix}
\end{equation}
where \(Y_0 = Z_0^{-1}\).  From reciprocity \(S = S^\top\) we find \(\mathbf{x} \equiv\mathbf{x}_2 = -\mathbf{x}_1\) and therefore
\begin{equation}
	S = 
	\begin{pmatrix}
	        r_{\upmu\upmu} & t_{1\upmu} & t_{2\upmu} & \dots \\
	        t_{1\upmu}     & r_{11}     & t_{21}     & \\
	        t_{2\upmu}     & t_{21}     & r_{22}     & \\
	        \vdots         &            &            & \ddots
	\end{pmatrix} =
	\begin{pmatrix}
            \frac{Y_0 - Y}{Y_0 + Y} & \frac{\sqrt{2Y_0}}{Y_0 + Y}\mathbf{x}^\top \\
            \frac{\sqrt{2Y_0}}{Y_0 + Y}\mathbf{x} & \trm{X} + \frac{1}{Y_0 + Y}\mathbf{x}\mathbf{x}^\top
        \end{pmatrix}.
        \label{eq:S-matrix}
\end{equation}
The first component of \(S\) connecting \(a_-\) and \(a_+\) is the reflection S-parameter \(S_{11}\) in the absence of reflections in the network (such as off a second transducer). The component of \(S\) connecting \(a_-\) to the SH0 coefficient \(a_\trm{b}\) is \(\tbmu\).  Its magnitude is conveniently expressed
\begin{equation}
	\left|\tbmu \right| = \sqrt{1 - \left|S_{11}^2\right|} \frac{\left|x_\trm{b}\right|}{\sqrt{2 G}}
\end{equation}where \(x_\trm{b}\) is the SH0 component of \(\mathbf{x}\) and is the coefficient \(a_\trm{b}\) for a \(1~\trm{V}\) drive. 
\(2G\left|V\right|^2\) is the total power dissipated by the transducer.

\section{De-embedding \(\tbmu\), \(\alpha\), and \(r_{\trm{bb}}\) from \(S_{21}\)}

\label{sec:deembed}
When making a transducer, especially one embedded in a network, \emph{e.g.}, a transducer coupled to a resonator, it is tempting to be satisfied with a well-matched \(S_{11}\).  Suppression of \(|S_{11}|\), \emph{i.e.}, low microwave reflections, seems to imply that microwaves are being converted to mechanical waves and that the device is efficient. Under this prescription, one would choose an IDT's width and simply tune its length until it is matched. This approach does not lead to efficient devices. 

\begin{figure}[h]
    \centering
    \includegraphics[width=1\columnwidth]{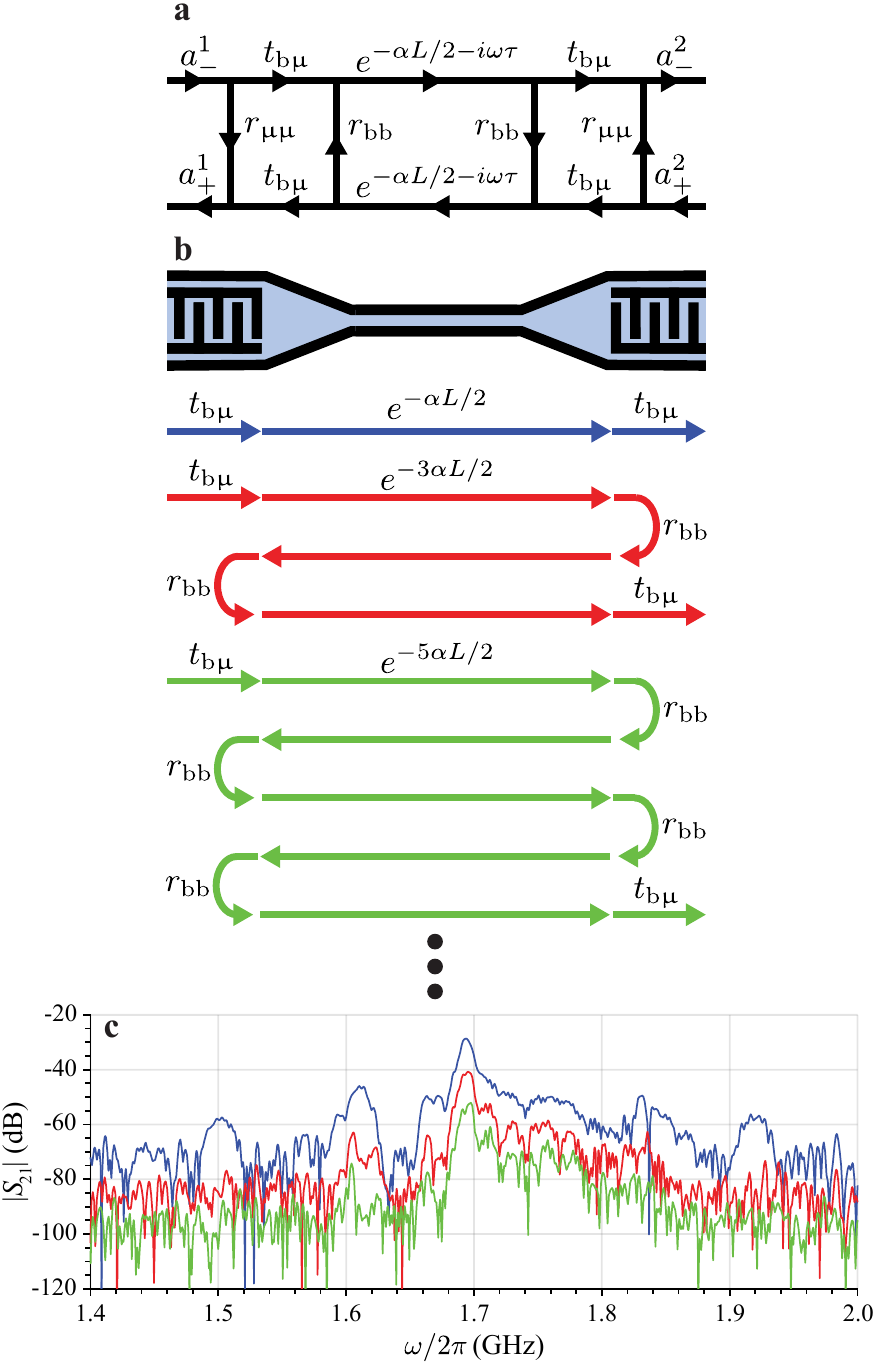}
    \caption{\textbf{a.} Signal-flow graph for a two-port device. \textbf{b.} Paths for the single, triple, and quintuple-transit (colored blue, red, and green respectively) corresponding to the \(S_{21}\) curves of matching color in \textbf{c.} \textbf{c.} \(S_{21}\) filtered by path.}
    \label{fig:signalFlow}
\end{figure}

A strong \(S_{11}\) dip is a \emph{necessary} but insufficient condition for efficiency (\(\left|\tbmu\right|^2 \rightarrow 1\)).  In a microwave or phononic network, reflections can strongly modify the response of a component.  Resonance can enhance the transmission through the device.  If network performance is the prime and only concern, measuring a resonator's intracavity phonon number against microwave input power, for example, will suffice.  But if the goal is to make a transducer which can serve as a general component, one that can be embedded in an arbitrary network and the response accurately predicted, we need to de-embed the transducer's response from the larger network response.

In Appendix~\ref{sec:solutions-to-S}, we describe how the full scattering matrix \(S\) can be computed by the FEM.  In Section~\ref{sec:IDT}, we show that transmission into the SH0 mode exceeds the total transmission into all other modes by \(10~\trm{dB}\).  This allows us to reduce the \(S\) matrix of Equation~\ref{eq:S-matrix} to two-ports
\begin{equation}
    S= 
        \begin{pmatrix}
             r_{\upmu\upmu} & \tbmu \\
             \tbmu & r_{\trm{bb}} 
        \end{pmatrix}.
\end{equation}
The S-matrix for the waveguide is
\begin{equation}
    S_\trm{wg} = 
        \begin{pmatrix}
             e^{-\alpha L/2 - i\omega\tau} & 0 \\
             0 & e^{-\alpha L/2 - i\omega\tau} 
        \end{pmatrix}
\end{equation}
where \(\tau = L/v_\trm{g}\) is the transit time of the waveguide.
The devices measured in Section~\ref{sec:measurements} consist of a transducer, waveguide, and transducer.  These components are cascaded in the signal flow graph in Figure~\ref{fig:signalFlow}a which can be reduced by standard methods~\cite{Pozar2012} to find
\begin{equation}
    S_{11} = r_{\upmu\upmu} + \frac{\tbmu^2 r_{\trm{bb}} e^{-\alpha L -2i\omega \tau}}{1 - r_\trm{bb}^2 e^{-\alpha L - 2i\omega\tau}}
\end{equation}
and 
\begin{equation}
    S_{21} = \frac{\tbmu^2 e^{-\alpha L /2 - i\omega\tau}}{1 - r_\trm{bb}^2e^{-\alpha L - 2i\omega\tau}}
\end{equation}
where ports 1 and 2 are the electrical port of the first and second transducer.
The second term in our expression for \(S_{11}\) comes from reflections \(r_\trm{bb}\) and gives rise to the Fabry-P\'erot peaks found on the blue side of \(\omega_0\) in Figure~\ref{fig:impulseResponse}c.

The impulse response \(h\prens{t}\) is computed by inverse Fourier transforming \(S_{21}\).
Expanding \(\prens{1 - r_\trm{bb}^2e^{-\alpha L - 2i\omega\tau}}^{-1}\) to \(\sum_n r_\trm{bb}^{2n}e^{-n\alpha L - 2in\omega\tau}\), each term represents an echo in the impulse response in Figure~\ref{fig:impulseResponse}e.
These echoes and the paths they take are diagrammed in Figure~\ref{fig:signalFlow}b.

Since the \(L = 800~\um\) device is long enough to resolve the echoes, the amplitudes of the echoes can be analyzed directly in the frequency domain by filtering out each echo in Figure~\ref{fig:impulseResponse}e associated with a path in Figure~\ref{fig:signalFlow}b and taking the Fourier transform.  The results of this procedure are inset to Figure~\ref{fig:impulseResponse}e but are reproduced larger below for clarity.   The transmission factor \(\tbmu^2\exp\prens{-\alpha L/2}\) is extracted from the first transit plotted in blue.

\section{Insertion loss from impedance mismatch}

\label{sec:2portMismatch}

In section~\ref{sec:measurements} we attribute a fraction of the insertion loss to impedance mismatch between the transducers and transmission lines. This mismatch in Figure~\ref{fig:impulseResponse}d is labeled the \emph{two-port mismatch}. Derived below, this quantity is the average of the ratios of the power dissipated by each transducer over the incident microwave power \(1-|S_{11}|^{2}/2-|S_{22}|^{2}/2\).

For any lossy, passive system
\begin{equation}
    |S_{11}|^{2}+|S_{12}|^{2}<1
    \label{eq:passive_con1}
\end{equation}
and
\begin{equation}
    |S_{22}|^{2}+|S_{21}|^{2}<1.
    \label{eq:passive_con2}
\end{equation}
Summing these conditions and assuming reciprocity, i.e. \(S_{21} = S_{12}\) we have
\begin{equation}
    |S_{11}|^{2}+|S_{22}|^{2}+2|S_{21}|^{2}<2
\end{equation}
Rearranging the above expression we get
\begin{equation}
    |S_{21}|^{2}<1-|S_{11}|^{2}/2-|S_{22}|^{2}/2
\end{equation}
The right-hand side which sets an upperbound on \(S_{21}\) is the two-port mismatch.

\end{document}